\documentclass[aip,rsi,amsmath,amssymb,reprint]{revtex4-1}
\pdfoutput=1
\usepackage{graphicx}
\usepackage{units}
\usepackage{dcolumn}
\usepackage{natbib} 

\begin{document}
\title{Adjustable coupling and \textit{in-situ} variable frequency EPR probe with loop-gap resonators for spectroscopy up to X-band}
\author{G. Joshi}
\affiliation{Department of Physics and Astronomy, Amherst College, Amherst, MA 01002, USA}
\author{J. Kubasek}
\affiliation{Department of Physics and Astronomy, Amherst College, Amherst, MA 01002, USA}
\author{I. Nikolov}
\affiliation{Department of Physics and Astronomy, Amherst College, Amherst, MA 01002, USA}
\author{B. Sheehan}
\affiliation{Department of Physics and Astronomy, Amherst College, Amherst, MA 01002, USA}
\affiliation{Department of Physics, University of Massachusetts
Amherst, MA 01003, USA}
\author{T. A. Costa}
\affiliation{Instituto de Quimica, Universidade Federal do Rio de Janeiro, Rio de Janeiro, RJ 21941-909, Brazil}
\author{R. A. All\~{a}o Cassaro}
\affiliation{Instituto de Quimica, Universidade Federal do Rio de Janeiro, Rio de Janeiro, RJ 21941-909, Brazil}
\author{Jonathan R. Friedman}
\email{jrfriedman@amherst.edu}
\affiliation{Department of Physics and Astronomy, Amherst College, Amherst, MA 01002, USA}
\date{\today}
\begin{abstract}
In standard electron paramagnetic resonance (EPR) spectroscopy, the frequency of an experiment is set and the spectrum is acquired using magnetic field as the independent variable.  There are cases in which it is desirable instead to fix the field and tune the frequency such as when studying avoided level crossings. We have designed and tested an adjustable frequency and variable coupling EPR probe with loop-gap resonators (LGRs) that works at a temperature down to 1.8 K. The frequency is tuned by adjusting the height of a dielectric piece of sapphire inserted into the gap of an LGR; coupling of the microwave antenna is varied with the height of antenna above the LGR. Both coupling antenna and dielectric are located within the cryogenic sample chamber, but their motion is controlled with external micrometers located outside the cryostat. The frequency of the LGR can be adjusted by more than 1 GHz. To cover a wide range of frequencies, different LGRs can be designed to cover frequencies up to X-band. We demonstrate the operation of our probe by mapping out avoided crossings for the Ni$_4$ single-molecule magnet to determine the tunnel splittings with high precision.   
\end{abstract}
\maketitle

\section{\label{sec:level1} Introduction \protect\\}
Electron paramagnetic resonance (EPR) spectroscopy is an indispensable tool in several branches of science as well as in industry. Several commercial spectrometers exist at certain well-defined frequencies to perform EPR spectroscopy. Several techniques are used for the resonant excitation of electronic spin to achieve magnetic resonance. Cavity resonators\cite{mettAxiallyUniformResonant2001,andersonCavitiesAxiallyUniform2002}, loop-gap resonators (LGR) \cite{hardySplitRingResonator1981,FRONCISZ1982515,woodLoopgapResonatorII1984} and planar circuit elements such as microstrip\cite{johanssonStriplineResonatorESR1974,cebulkaSubKelvin100MK2019} and coplanar waveguide resonators (CPW) \cite{claussOptimizationCoplanarWaveguide2015,malissaSuperconductingCoplanarWaveguide2013} are widely used for the excitation of spins. Most of the commercial EPR spectrometers are built with cavity resonators with predefined frequencies. Even though such resonators provide reasonably good homogeneity of  radio frequency (RF) magnetic field (\textit{B$_1$})  over a large volume of sample, their construction below X band is challenging because of the large size of resonator cavities needed \cite{fukudaDevelopmentMultifrequencyESR2016}. LGRs and planar resonators are frequently used for EPR spectroscopy to cover a wide range of frequencies without the need of a large resonator size. LGRs typically have high quality factors ($Q >2000$), good RF magnetic field (\textit{B$_1$}) homogeneity in the sample space,  and a high  filling factor, making them highly sensitive resonators\cite{blankRecentTrendsHigh2017}. However, their structures and fabrication becomes difficult when the frequency increases beyond X-band\cite{sidabrasMultipurposeEPRLoopgap2007,denysenkovQBandLoopGapResonator2017}. On the other hand, planar resonators have usually lower $Q$ ($<100$) and can be fabricated to cover a wide frequency range. Indeed, higher harmonics of the CPW resonators have been successfully used for excitation of spins in thin-film devices\cite{joshiSeparatingHyperfineSpinorbit2016} to obtain multifrequency magnetic resonance spectra without changing the resonator.  

When a system's energy levels have a non-trivial dependence on magnetic field, such as when energy levels have avoided crossings, it is desirable to tune the frequency of the resonator instead of varying the field to change the level structure.  Such cases arise in molecular nanomagnets (MNMs) where the ``tunnel splitting" of an  avoided  crossing can be mapped out by varying both the excitation frequency and magnetic field.     Here we  report the design and implementation of an EPR resonator probe using LGRs that allows the \textit{in situ} tuning of frequency by $>\unit[1]{GHz}$ while also allowing the coupling of antenna and resonator to be optimized at each frequency. We have successfully used this novel probe to determine the tunnel splitting gaps of a the Ni$_4$ MNM with high precision.  

MNMs are  metal-organic compounds with a ground-state magnetic moment.  Typically, an MNM has $S >\frac{1}{2}$ and   significant magnetic anisotropy.  The chemical engineering of the molecules permits the tuning of their  properties thus making them attractive as spin qubits\cite{wedgeChemicalEngineeringMolecular2012}. Recent studies of atomic clock transitions, a phenomenon marked by a significant increase in the coherence time ($T_2$) of spin states at an avoided level crossing, in spin systems\cite{wolfowiczAtomicClockTransitions2013, shiddiqEnhancingCoherenceMolecular2016, collettClockTransitionCr7Mn2019} increase the viability of such systems as qubits. Experiments at atomic clock transitions require precise frequency control to determine the precise values of tunnel splitting and to study coherence in the vicinity of the clock transition. The probe we designed and constructed allows for the precise frequency control needed to map out an avoided crossing through EPR. The details of design and implementation of the probe are  described in the following section.

\section{Design and Implementation}
\begin{figure}[ht]
\centering
\includegraphics{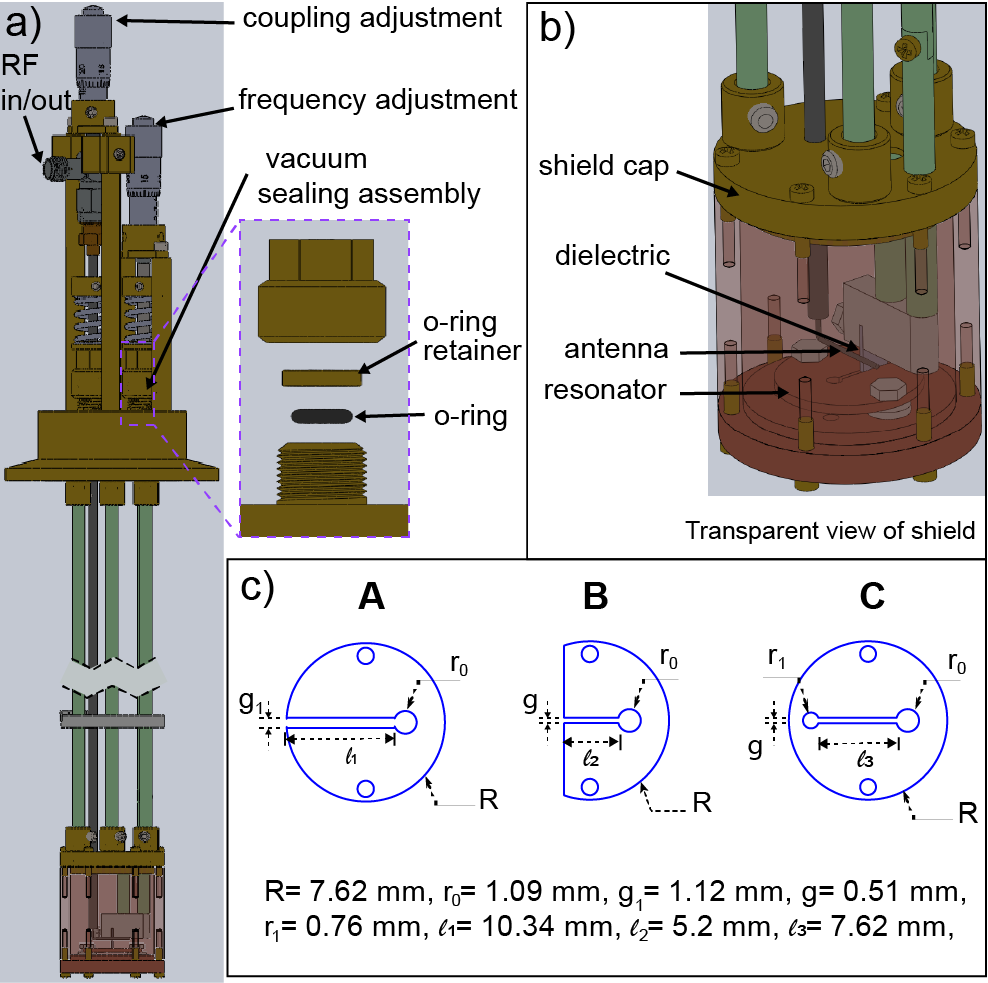}
\caption{Design of resonator probe with tunable frequency and coupling. (a) CAD model of the resonator probe compatible with the Quantum Design PPMS sample chamber. The vacuum sealing assembly that allows motion of signal cable/control rod is shown in the zoomed outset.(b) Close up view of the interior of the shield, showing LGR, antenna, dielectric and the position control mechanism. The height of the dielectric inside the gap of LGR determines the resonant frequency while the coupling antenna height above the resonator surface determines coupling of the radiation to the resonator. (c)  Scaled drawing of the resonators used to collect data for this article. }
\label{fig1}
\end{figure}

We have designed and implemented a cryogenic EPR probe where the resonant frequency of a loop-gap resonator (LGR) can be varied \textit{in-situ} over a $>$\unit[1]{GHz} range. In tandem, the coupling can also be adjusted to achieve optimal experimental conditions at all frequencies. The heart of the apparatus consists of the LGR, which can be oriented such that RF magnetic field is either parallel or perpendicular to the applied DC field.  A sample is positioned in the loop of the LGR, which is then mounted on the base of shield with nylon washers and screws (cf. Fig.~\ref{fig1}(c)) to electrically isolate the LGR. The resonant frequency of the LGR is determined by the inductance of the loop and capacitance of the gap\cite{FRONCISZ1982515}.  By inserting a low-loss dielectric into the gap, the capacitance is increased, lowering the resonant frequency. 

In our apparatus, the loop-gap resonator is enclosed within a cylindrical copper radiation shield that prevents radiative losses. The size of the shield used in our probe is  about \unit[39]{mm} outer diameter, \unit[20]{mm} long, \unit[2.8]{mm} thick wall and \unit[4.4]{mm} thick bottom and top plates. Radiation is coupled to the resonator through an antenna formed by stripping the outer conductor and dielectric off the last few millimeters of the copper coaxial cable and bending the remaining inner conductor so that it is parallel to the plane of the resonator.  The unique character of our apparatus is that we are able to control the position of the dielectric within the LGR and the height of the antenna above the resonator.  To achieve this we built two separate translational motion controllers, one for a dielectric piece that moves through the gap of the LGR, and the other for the signal cable connected to the RF antenna, thus adjusting the coupling strength.  The two controllers are located outside the sample chamber on the top of probe, as shown in Fig.~\ref{fig1}(a), and thus permit adjustments while the resonator and sample are at cryogenic temperatures. The dielectric holder is connected to the external micrometer screw (\unit[0.5]{mm/turn}) by a rod running the length of the probe. The top portion of the rod (the section that passes through the vacuum sealing assembly -- see below) is made out of stainless steel.  This is joined to a G10 rod that runs into the shield and attaches to the dielectric holder. It contains a slot into which the dielectric is inserted.  A nylon screw is used to affix the dielectric in place.  The dielectric holder is bolted onto the G10 rod. To prevent the rotational motion of dielectric holder when it is moved, the holder is also attached to another rod, parallel to the first.  The two rods move up and down through two holes in the top of shield, preventing any rotation. This keeps the dielectric slab parallel to the gap of the LGR, as shown in the Fig.~\ref{fig1}(b). The figure shows the resonator oriented with its loop axis parallel to the magnetic field of the superconducting magnet (parallel mode). To use the apparatus in perpendicular mode, the resonator \textbf{B} is placed in a delrin supporting frame (not shown) such that the gap end of resonator edge lies upward and the slot of the dielectric holder is perpendicular to the plane of resonator. The same dielectric holder is used to change the  frequency but the coupling is maintained by adjusting the the height of antenna relative to the flat edge of resonator \textbf{B}. To test the working of probe in perpendicular mode, an EPR standard sample, DPPH (2,2-diphenyl-1-picrylhydrazyl), was placed in the loop at room temperature and the magnetic field swept. The expected EPR transitions were observed at \textit{g} $\approx$ 2 (spectra not shown). 

A semi-rigid coaxial cable carries signals to/from the resonator.  The cable consists of two parts; the upper part, running most of the length of the probe has a BeCu inner conductor and a stainless steel outer conductor.  A few inches above the shield, the upper cable connects to a short section of copper coax, which passes through a hole in the shield and terminates in the antenna.  This is done to keep the magnetic stainless steel away from the LGR and sample.  Although the cable enters the shield through a hole in the top plate, remarkably no significant reduction in $Q$ of LGR is found from this ``breaking" of the shield's Faraday cage. The coaxial cable is connected to an external microwave source via a right-angled SMA adapter. Another micrometer, with the same specifications as the first, pushes on the SMA adapter to adjust the height of the antenna relative to the resonator in order to achieve the desired coupling.  A spring under each micrometer provides sufficient tension to allow the rod or cable to rise when the micrometer is retracted. The apparatus contains only one coaxial cable and thus operates in reflection mode with spectra obtained from the S$_{11}$ coefficient. 


In order for the apparatus to operate at cryogenic temperatures the probe must be leak tight while in the sample chamber. We also need to ensure vacuum integrity under translational motion of the rod and coax cable that pass through the flange.  To achieve this, we limited the number of vacuum seals.  The top of the probe is based on a standard brass KF-40 flange that fits to the top of the PPMS sample chamber.  We created two custom feedthroughs through the flange.  The flange was monolithically machined (using a CNC milling machine) to create two raised feedthrough bases, each with  $\nicefrac{5}{16}-32$ external screw threads and a axial hole for the rod/coax to pass through.   The vacuum seal is created using an o-ring and o-ring retaining ring, as shown in exploded view in Fig.~\ref{fig1}(a) (outset). By hand-tightening a nut on the top of the assembly, the o-ring is compressed against the rod or cable that passes through it, making the seal.  In order to make a dynamic seal that is robust under translation of the rod (cable), the rod (cable) is polished to a shine, producing a smooth surface to which the o-ring mates reliably at all times.

The flange is connected to the top of shield by three G10 rods of length of $\approx$\unit[1]{m}, such that the center of shield is at the center of homogeneity for the superconducting magnet of the PPMS. To attach the G10 rods, three short uprights are attached on the inner side of KF 40 blank and on the top of shield. 3D-printed circular discs of ABS plastic are affixed to the G10 rods at several places along the probe.  These serve to provide rigidity to the probe and also act as baffles to mitigate convective heat transfer along the length of the probe.  

We have used three different LGRs to obtain data for this manuscript. Scaled drawing of the resonators, labelled \textbf{A}, \textbf{B}, and \textbf{C}, are shown in Fig.~\ref{fig1}(c). The resonant frequencies of the resonators without dielectric are $f_\textbf{A}=$ \unit[4.50]{GHz}, $f_\textbf{B}=$ \unit[5.17]{GHz} and $f_\textbf{C}=$ \unit[6.51]{GHz}. Most dimensions of the resonators are indicated in the Fig.~\ref{fig1}(c).  The thickness of resonators \textbf{A} and \textbf{B} is \unit[3]{mm}, while \textbf{C} is \unit[2]{mm}. The two loops in resonator \textbf{C} are responsible for its higher resonant frequency because of the parallel combination of the two inductances of the LGR. For each resonator, the frequency is lowered by insertion of a dielectric, which is typically a small slab of sapphire.  

\begin{figure}[ht]
\centering
\includegraphics{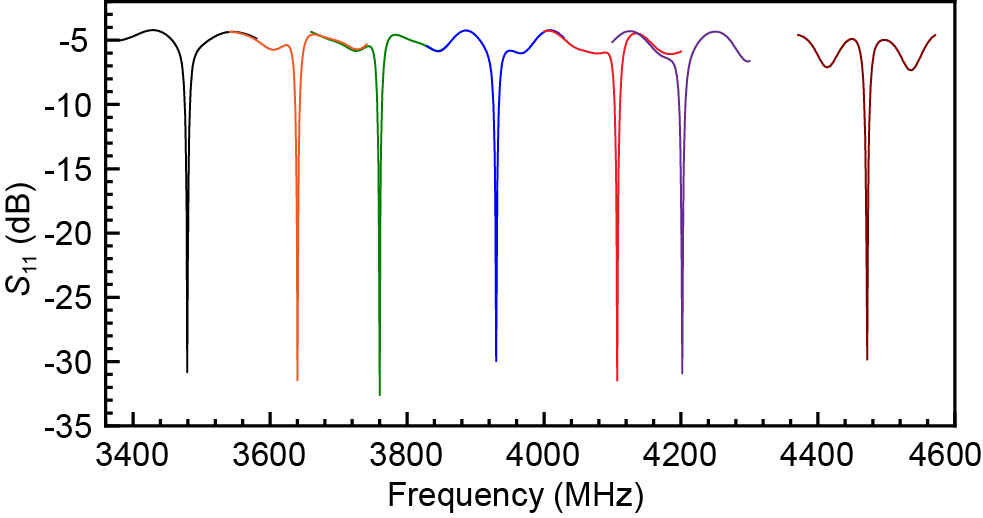}
\caption{Resonant response of LGR obtained at \unit[1.8]{K} at various positions of the dielectric (sapphire) in the gap of resonator, adjusted using a micrometer screw outside of probe. Near critical coupling at each frequency is achieved by adjusting the height of the antenna from LGR surface. The resonator used is illustrated as resonator \textbf{A} (Fig.~\ref{fig1}(c)).}
\label{fig2}    
\end{figure}

To demonstrate the \textit{in-situ} variation of frequency and adjustment of coupling at cryogenic temperatures resonator \textbf{A} was used. A vector network analyzer (Keysight E5063A) was used to characterize the response of the resonator (\textit{S}$_{11}$) for each position of the dielectric in the LGR gap. Figure~\ref{fig2} shows a selected set of \textit{S}$_{11}$ vs. frequency curves taken at \unit[1.8]{K} for several different positions of a  sapphire slab of size $\approx$ \unit[3]{mm}$\times$ \unit[4]{mm} with \unit[1]{mm} thickness. For each position of the dielectric, the antenna position was adjusted to achieve near critical coupling to the LGR.  
The \textit{Q} factor of the resonators at \unit[10]{K} is $>$ 2000 making them well suited for continuous wave (cw) EPR spectroscopy. To perform pulsed EPR experiments, a lower $Q$ is typically required.  $Q$ can be reduced by overcoupling the antenna to the resonator at the cost of higher reflection at the resonant frequency. We have also reduced $Q$ by placing a small piece of microwave absorbing material (Eccosorb LS-26) at the bottom of shield just below the resonator. The resonant spectra shown in the Fig.~\ref{fig2} are taken when Q-factor is reduced to $\approx$ 500 using this latter technique. 
\section{Testing the apparatus}
\begin{figure}[ht]
\centering
\includegraphics{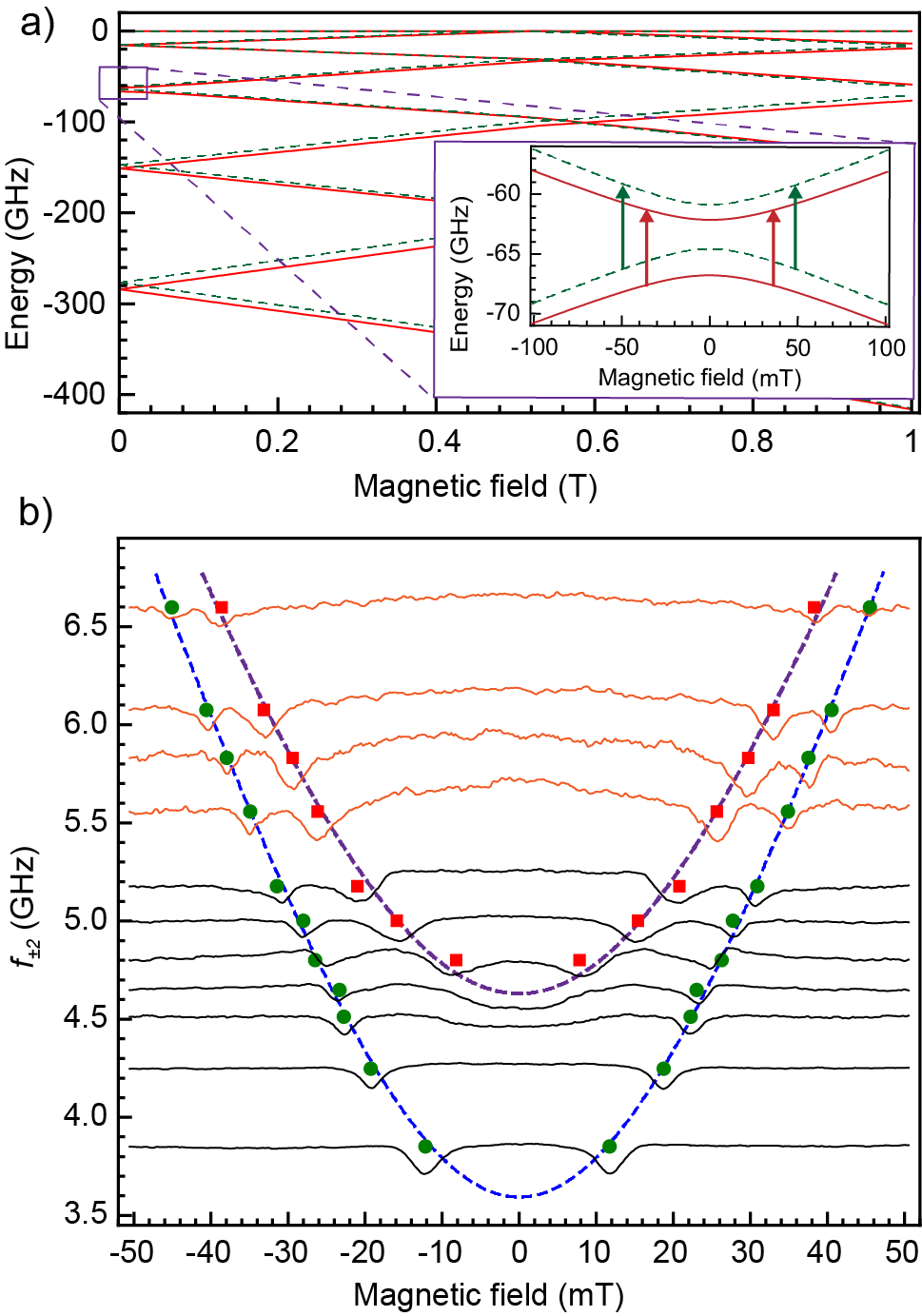}
\caption{Measurement of the tunnel splitting gap in Ni$_4$ molecular magnet using the instrument. (a) Energy-level diagram of Ni$_4$  when the  magnetic field is  along the easy axis. We study the transitions near the avoided level crossing of the $m=\pm 2$ states, as shown in the inset of (a). The red solid and dashed green energy levels represents the two different conformational states of molecule at low temperatures. (b) cw EPR spectra measured at various frequencies using resonator \textbf{B} (black curves) and \textbf{C} (orange); cf.~Fig~\ref{fig1}(c). EPR peak positions for the two conformational states are indicated by green circles and red squares. Fitting the peak positions (dashed curves) to Eq.~\ref{fit} yield the tunnel splitting for each conformational state. }
\label{fig3}
\end{figure}

In order to test the novel probe, we chose a well studied spin \textit{S}= 4 single molecule magnet [Ni(hmp)(dmb)Cl]$_4$, hereafter called `Ni$_4$', where hmp stands for 2-hydroxymethylpyridine and dmb for 3,3-dimethyl-1-butanol \cite{yangFastMagnetizationTunneling2006a}. When the field is applied along the easy (z) axis of magnetization for the molecule, the spin Hamiltonian of Ni$_4$ is well described by\cite{collettPrecisionESRMeasurements2016}, 
\begin{equation}
\mathcal{H}=-D{S_Z}^2-A{S_z}^4 + g_z \mu_B B_zS_z+\mathcal{H}',
\end{equation}
where \textit{D} and \textit{A} are the axial anisotropy parameters, $g_z$ is the $g$ factor, and $B_z$ is the applied magnetic field component along the $z$ axis. The term  $\mathcal{H}'$ does not commute with $S_z$ and hence is responsible for the occurrence of tunneling\cite{friedmanSingleMoleculeNanomagnets2010}. For Ni$_4$, \textit{D} $>$ 0 and \textit{A} $>$ 0 so the system has lowest energy when its moment is parallel or antiparallel to the $z$ axis. If $\mathcal{H}'$ = 0, the angular momentum component $S_z$ is conserved and the energy states are described in terms of quantum numbers \textit{m}. Fig.~\ref{fig3}(a) shows the system's energy eigenvalues as a function of magnetic field along easy axis of Ni$_4$.
The dashed green and solid red lines correspond to two different conformational states (isomers) of the  Ni$_4$ molecule, each of which have slightly different anisotropy parameters.  For this molecule, $\mathcal{H}'$ contains fourth-order transverse anisotropy terms:
\begin{equation}
    \mathcal{H}'=C(S^4_+ +S^4_-).
\end{equation}
$\mathcal{H}'$ breaks the rotational symmetry of the Hamiltonian and lifts the degeneracy between levels at zero field (as well as other fields where levels would otherwise cross). Instead, nearly degenerate levels form an avoided crossing with a minimum energy gap called the tunnel splitting $\Delta$.  For the avoided crossing between superpositions of  $m=\pm 2$ states, $\Delta_{\pm2} \approx$ \unit[4]{GHz} at zero field. The inset of Fig.~\ref{fig3}(a) highlights the tunnel splitting gap for these states near zero field. The slight difference in the parameters of the two conformational states leads to the doubling of the EPR spectra\cite{lawrenceDisorderIntermolecularInteractions2008,chenObservationTunnelingAssistedHighly2016} provided that the microwave frequency is larger than tunneling gaps for both conformational states. The green and red arrows in the Fig.~\ref{fig3}(a) inset shows the two transitions corresponding to the two different conformational states at a particular excitation frequency. These transitions are forbidden in perpendicular mode EPR, requiring that they need to be measured in parallel mode.  The resonant frequency near an avoided crossing centered at field $B_c$ can be described by\cite{collettPrecisionESRMeasurements2016}
\begin{equation}
f_{m,m^\prime}= \sqrt{\Delta^2_{m,m^\prime} + [g_z\mu_B (m-m^\prime)(B-B_c)]^2},\label{fit}
\end{equation}
where $\Delta_{m,m^\prime}$ is the minimum tunnelling gap at $B = B_c$, and $m$ and $m^\prime$ are the quantum numbers associated with the states in the avoided crossing. 

In this experiment, we used the probe described herein for direct measurement of the zero-field ($B_c=0$) tunnel splitting parameters of a Ni$_4$ SMM sample for the states $m,m^\prime=\pm 2$. The sample was produced by co-crystallization of Ni$_4$ with a diamagnetic analogue Zn$_4$ to produce a \unit[5]{\%} dilute sample, as described in previous work\cite{collettPrecisionESRMeasurements2016}. In that work, the tunnel splitting parameters of Ni$_4$ were determined by using multi-frequency EPR methods near the zero-field splitting of state $\unit[m = \pm]{2}$, requiring the apparatus to be thermally cycled every time the frequency needed to be changed.   With the new probe we have been able to measure a broad range of spectra in a single cooling of sample, changing the frequency in situ.  Furthermore, in previous work, the antenna coupling was adjusted at room temperature but would change as the system cooled, meaning that the experiment rarely operated near critical coupling.  Using the new apparatus, we were able to circumvent these difficulties and efficiently obtain multiple spectra with near critical coupling and thereby precisely determine the zero-field splitting parameters of Ni$_4$. 

A \unit[5]{\%} Ni$_4$  single crystal was placed at the center of loop of LGR and oriented with its easy axis parallel to the loop/field axis.  The Q-factor of the resonator was measured as a function of magnetic field at \unit[10]{K}. Sweeping the magnetic field induces eddy currents in the shield and other metallic parts of the apparatus.  These, in turn, produce a field that partially cancels the applied field, producing an apparent shift of our spectra.  Sweeping the field in the opposite direction (down vs. up) produces an opposite shift and sweeping faster produces a larger shift, confirming the inductive nature of the shifts.  Subtracting the shift from each spectrum, resulted in spectra that were all nicely symmetric about zero field. The black (orange) spectra shown in the Fig.~\ref{fig3}(b) were obtained using resonators \textbf{B}  (\textbf{C}) shown in Fig.~\ref{fig1}(c). For each resonator, the frequency was tuned over more than a \unit[1]{GHz} range in a single cooling cycle of apparatus. The spectra plotted in the Fig.~\ref{fig3}(b) have been shifted vertically so that the baseline of each spectrum aligns with the resonant frequency indicated in the plot.

To find the resonant position, Lorentzian functions are fit to each spectral peak to determine the center of each EPR transition. In Fig.~\ref{fig3}(b) the peak position of one conformational state are represented by the green circles and those of the other conformational state are represented by red squares. These data points are fit to Eq.~\ref{fit} with $B_c=0$, $m=2$ and $m^\prime=-2$ to determine the tunnel splitting for each molecular conformation. We found that the tunneling gaps for the two conformations states are $\Delta^{(1)}_{\pm 2} = \unit[4.63(3)]{GHz} $ and $\Delta^{(2)}_{\pm 2} = \unit[3.59(2)]{GHz}$. These values are in excellent agreement with previous measurements\cite{collettPrecisionESRMeasurements2016} with the latter value now being determined with an order of magnitude higher precision.

\section{Conclusion}

The apparatus described above is highly versatile, permitting frequency tuning and coupling optimization while the sample and resonator are at cryogenic temperatures, making it easy to reliably obtain multiple spectra in a single cool down.  This increases efficiency for data acquisition.  In addition, being able to tune the frequency \textit{in situ} may be a substantial improvement when working with air sensitive samples or materials that do not tolerate repeated thermal cycling.  We typically use sapphire as a dielectic for frequency tuning.  With a dielectric constant of 9.3 - 11.5, this is sufficient for most of our work.  However, one could substitute another dielectric with larger permittivity (such as rutile) to obtain a larger range of frequency tunability.  

\begin{acknowledgments}
We thank C. Collett and G. Chen for useful conversations and advice.  We are indebted to R. Winn for 3D printing of portions of the apparatus. Support for this work was provided by the National Science Foundation under Grant No.~DMR-1708692, and by the Amherst College Dean of Faculty. R. A. A. Cassaro thanks FAPERJ and CNPq for the financial support. T. A. Costa Acknowledges CNPq for the fellowship.
\end{acknowledgments}

\bibliography{ms.bbl}

\begin{thebibliography}{23}%
\makeatletter
\providecommand \@ifxundefined [1]{%
 \@ifx{#1\undefined}
}%
\providecommand \@ifnum [1]{%
 \ifnum #1\expandafter \@firstoftwo
 \else \expandafter \@secondoftwo
 \fi
}%
\providecommand \@ifx [1]{%
 \ifx #1\expandafter \@firstoftwo
 \else \expandafter \@secondoftwo
 \fi
}%
\providecommand \natexlab [1]{#1}%
\providecommand \enquote  [1]{``#1''}%
\providecommand \bibnamefont  [1]{#1}%
\providecommand \bibfnamefont [1]{#1}%
\providecommand \citenamefont [1]{#1}%
\providecommand \href@noop [0]{\@secondoftwo}%
\providecommand \href [0]{\begingroup \@sanitize@url \@href}%
\providecommand \@href[1]{\@@startlink{#1}\@@href}%
\providecommand \@@href[1]{\endgroup#1\@@endlink}%
\providecommand \@sanitize@url [0]{\catcode `\\12\catcode `\$12\catcode
  `\&12\catcode `\#12\catcode `\^12\catcode `\_12\catcode `\%12\relax}%
\providecommand \@@startlink[1]{}%
\providecommand \@@endlink[0]{}%
\providecommand \url  [0]{\begingroup\@sanitize@url \@url }%
\providecommand \@url [1]{\endgroup\@href {#1}{\urlprefix }}%
\providecommand \urlprefix  [0]{URL }%
\providecommand \Eprint [0]{\href }%
\providecommand \doibase [0]{http://dx.doi.org/}%
\providecommand \selectlanguage [0]{\@gobble}%
\providecommand \bibinfo  [0]{\@secondoftwo}%
\providecommand \bibfield  [0]{\@secondoftwo}%
\providecommand \translation [1]{[#1]}%
\providecommand \BibitemOpen [0]{}%
\providecommand \bibitemStop [0]{}%
\providecommand \bibitemNoStop [0]{.\EOS\space}%
\providecommand \EOS [0]{\spacefactor3000\relax}%
\providecommand \BibitemShut  [1]{\csname bibitem#1\endcsname}%
\let\auto@bib@innerbib\@empty
\bibitem [{\citenamefont {Mett}, \citenamefont {Froncisz},\ and\ \citenamefont
  {Hyde}(2001)}]{mettAxiallyUniformResonant2001}%
  \BibitemOpen
  \bibfield  {author} {\bibinfo {author} {\bibfnamefont {R.~R.}\ \bibnamefont
  {Mett}}, \bibinfo {author} {\bibfnamefont {W.}~\bibnamefont {Froncisz}}, \
  and\ \bibinfo {author} {\bibfnamefont {J.~S.}\ \bibnamefont {Hyde}},\ }\href
  {\doibase 10.1063/1.1405796} {\bibfield  {journal} {\bibinfo  {journal}
  {Review of Scientific Instruments}\ }\textbf {\bibinfo {volume} {72}},\
  \bibinfo {pages} {4188} (\bibinfo {year} {2001})}\BibitemShut {NoStop}%
\bibitem [{\citenamefont {Anderson}, \citenamefont {Mett},\ and\ \citenamefont
  {Hyde}(2002)}]{andersonCavitiesAxiallyUniform2002}%
  \BibitemOpen
  \bibfield  {author} {\bibinfo {author} {\bibfnamefont {J.~R.}\ \bibnamefont
  {Anderson}}, \bibinfo {author} {\bibfnamefont {R.~R.}\ \bibnamefont {Mett}},
  \ and\ \bibinfo {author} {\bibfnamefont {J.~S.}\ \bibnamefont {Hyde}},\
  }\href {\doibase 10.1063/1.1491032} {\bibfield  {journal} {\bibinfo
  {journal} {Review of Scientific Instruments}\ }\textbf {\bibinfo {volume}
  {73}},\ \bibinfo {pages} {3027} (\bibinfo {year} {2002})}\BibitemShut
  {NoStop}%
\bibitem [{\citenamefont {Hardy}\ and\ \citenamefont
  {Whitehead}(1981)}]{hardySplitRingResonator1981}%
  \BibitemOpen
  \bibfield  {author} {\bibinfo {author} {\bibfnamefont {W.~N.}\ \bibnamefont
  {Hardy}}\ and\ \bibinfo {author} {\bibfnamefont {L.~A.}\ \bibnamefont
  {Whitehead}},\ }\href {\doibase 10.1063/1.1136574} {\bibfield  {journal}
  {\bibinfo  {journal} {Review of Scientific Instruments}\ }\textbf {\bibinfo
  {volume} {52}},\ \bibinfo {pages} {213} (\bibinfo {year} {1981})}\BibitemShut
  {NoStop}%
\bibitem [{\citenamefont {Froncisz}\ and\ \citenamefont
  {Hyde}(1982)}]{FRONCISZ1982515}%
  \BibitemOpen
  \bibfield  {author} {\bibinfo {author} {\bibfnamefont {W.}~\bibnamefont
  {Froncisz}}\ and\ \bibinfo {author} {\bibfnamefont {J.~S.}\ \bibnamefont
  {Hyde}},\ }\href {\doibase https://doi.org/10.1016/0022-2364(82)90221-9}
  {\bibfield  {journal} {\bibinfo  {journal} {Journal of Magnetic Resonance
  (1969)}\ }\textbf {\bibinfo {volume} {47}},\ \bibinfo {pages} {515} (\bibinfo
  {year} {1982})}\BibitemShut {NoStop}%
\bibitem [{\citenamefont {Wood}, \citenamefont {Froncisz},\ and\ \citenamefont
  {Hyde}(1984)}]{woodLoopgapResonatorII1984}%
  \BibitemOpen
  \bibfield  {author} {\bibinfo {author} {\bibfnamefont {R.~L.}\ \bibnamefont
  {Wood}}, \bibinfo {author} {\bibfnamefont {W.}~\bibnamefont {Froncisz}}, \
  and\ \bibinfo {author} {\bibfnamefont {J.~S.}\ \bibnamefont {Hyde}},\ }\href
  {\doibase 10.1016/0022-2364(84)90214-2} {\bibfield  {journal} {\bibinfo
  {journal} {Journal of Magnetic Resonance (1969)}\ }\textbf {\bibinfo {volume}
  {58}},\ \bibinfo {pages} {243} (\bibinfo {year} {1984})}\BibitemShut
  {NoStop}%
\bibitem [{\citenamefont {Johansson}\ \emph {et~al.}(1974)\citenamefont
  {Johansson}, \citenamefont {Haraldson}, \citenamefont {Pettersson},\ and\
  \citenamefont {Beckman}}]{johanssonStriplineResonatorESR1974}%
  \BibitemOpen
  \bibfield  {author} {\bibinfo {author} {\bibfnamefont {B.}~\bibnamefont
  {Johansson}}, \bibinfo {author} {\bibfnamefont {S.}~\bibnamefont
  {Haraldson}}, \bibinfo {author} {\bibfnamefont {L.}~\bibnamefont
  {Pettersson}}, \ and\ \bibinfo {author} {\bibfnamefont {O.}~\bibnamefont
  {Beckman}},\ }\href {\doibase 10.1063/1.1686524} {\bibfield  {journal}
  {\bibinfo  {journal} {Review of Scientific Instruments}\ }\textbf {\bibinfo
  {volume} {45}},\ \bibinfo {pages} {1445} (\bibinfo {year}
  {1974})}\BibitemShut {NoStop}%
\bibitem [{\citenamefont {Cebulka}\ and\ \citenamefont
  {Del~Barco}(2019)}]{cebulkaSubKelvin100MK2019}%
  \BibitemOpen
  \bibfield  {author} {\bibinfo {author} {\bibfnamefont {R.}~\bibnamefont
  {Cebulka}}\ and\ \bibinfo {author} {\bibfnamefont {E.}~\bibnamefont
  {Del~Barco}},\ }\href {\doibase 10.1063/1.5097563} {\bibfield  {journal}
  {\bibinfo  {journal} {Review of Scientific Instruments}\ }\textbf {\bibinfo
  {volume} {90}},\ \bibinfo {pages} {085106} (\bibinfo {year}
  {2019})}\BibitemShut {NoStop}%
\bibitem [{\citenamefont {Clauss}, \citenamefont {Dressel},\ and\ \citenamefont
  {Scheffler}(2015)}]{claussOptimizationCoplanarWaveguide2015}%
  \BibitemOpen
  \bibfield  {author} {\bibinfo {author} {\bibfnamefont {C.}~\bibnamefont
  {Clauss}}, \bibinfo {author} {\bibfnamefont {M.}~\bibnamefont {Dressel}}, \
  and\ \bibinfo {author} {\bibfnamefont {M.}~\bibnamefont {Scheffler}},\ }\href
  {\doibase 10.1088/1742-6596/592/1/012146} {\bibfield  {journal} {\bibinfo
  {journal} {Journal of Physics: Conference Series}\ }\textbf {\bibinfo
  {volume} {592}},\ \bibinfo {pages} {012146} (\bibinfo {year}
  {2015})}\BibitemShut {NoStop}%
\bibitem [{\citenamefont {Malissa}\ \emph {et~al.}(2013)\citenamefont
  {Malissa}, \citenamefont {Schuster}, \citenamefont {Tyryshkin}, \citenamefont
  {Houck},\ and\ \citenamefont
  {Lyon}}]{malissaSuperconductingCoplanarWaveguide2013}%
  \BibitemOpen
  \bibfield  {author} {\bibinfo {author} {\bibfnamefont {H.}~\bibnamefont
  {Malissa}}, \bibinfo {author} {\bibfnamefont {D.~I.}\ \bibnamefont
  {Schuster}}, \bibinfo {author} {\bibfnamefont {A.~M.}\ \bibnamefont
  {Tyryshkin}}, \bibinfo {author} {\bibfnamefont {A.~A.}\ \bibnamefont
  {Houck}}, \ and\ \bibinfo {author} {\bibfnamefont {S.~A.}\ \bibnamefont
  {Lyon}},\ }\href {\doibase 10.1063/1.4792205} {\bibfield  {journal} {\bibinfo
   {journal} {Review of Scientific Instruments}\ }\textbf {\bibinfo {volume}
  {84}},\ \bibinfo {pages} {025116} (\bibinfo {year} {2013})}\BibitemShut
  {NoStop}%
\bibitem [{\citenamefont {Fukuda}\ and\ \citenamefont
  {Asakawa}(2016)}]{fukudaDevelopmentMultifrequencyESR2016}%
  \BibitemOpen
  \bibfield  {author} {\bibinfo {author} {\bibfnamefont {K.}~\bibnamefont
  {Fukuda}}\ and\ \bibinfo {author} {\bibfnamefont {N.}~\bibnamefont
  {Asakawa}},\ }\href {\doibase 10.1063/1.4967712} {\bibfield  {journal}
  {\bibinfo  {journal} {Review of Scientific Instruments}\ }\textbf {\bibinfo
  {volume} {87}},\ \bibinfo {pages} {113106} (\bibinfo {year}
  {2016})}\BibitemShut {NoStop}%
\bibitem [{\citenamefont {Blank}, \citenamefont {Twig},\ and\ \citenamefont
  {Ishay}(2017)}]{blankRecentTrendsHigh2017}%
  \BibitemOpen
  \bibfield  {author} {\bibinfo {author} {\bibfnamefont {A.}~\bibnamefont
  {Blank}}, \bibinfo {author} {\bibfnamefont {Y.}~\bibnamefont {Twig}}, \ and\
  \bibinfo {author} {\bibfnamefont {Y.}~\bibnamefont {Ishay}},\ }\href
  {\doibase 10.1016/j.jmr.2017.02.019} {\bibfield  {journal} {\bibinfo
  {journal} {Journal of Magnetic Resonance}\ }\textbf {\bibinfo {volume}
  {280}},\ \bibinfo {pages} {20} (\bibinfo {year} {2017})}\BibitemShut
  {NoStop}%
\bibitem [{\citenamefont {Sidabras}\ \emph {et~al.}(2007)\citenamefont
  {Sidabras}, \citenamefont {Mett}, \citenamefont {Froncisz}, \citenamefont
  {Camenisch}, \citenamefont {Anderson},\ and\ \citenamefont
  {Hyde}}]{sidabrasMultipurposeEPRLoopgap2007}%
  \BibitemOpen
  \bibfield  {author} {\bibinfo {author} {\bibfnamefont {J.~W.}\ \bibnamefont
  {Sidabras}}, \bibinfo {author} {\bibfnamefont {R.~R.}\ \bibnamefont {Mett}},
  \bibinfo {author} {\bibfnamefont {W.}~\bibnamefont {Froncisz}}, \bibinfo
  {author} {\bibfnamefont {T.~G.}\ \bibnamefont {Camenisch}}, \bibinfo {author}
  {\bibfnamefont {J.~R.}\ \bibnamefont {Anderson}}, \ and\ \bibinfo {author}
  {\bibfnamefont {J.~S.}\ \bibnamefont {Hyde}},\ }\href {\doibase
  10.1063/1.2709746} {\bibfield  {journal} {\bibinfo  {journal} {Review of
  Scientific Instruments}\ }\textbf {\bibinfo {volume} {78}},\ \bibinfo {pages}
  {034701} (\bibinfo {year} {2007})}\BibitemShut {NoStop}%
\bibitem [{\citenamefont {Denysenkov}, \citenamefont {{van Os}},\ and\
  \citenamefont {Prisner}(2017)}]{denysenkovQBandLoopGapResonator2017}%
  \BibitemOpen
  \bibfield  {author} {\bibinfo {author} {\bibfnamefont {V.}~\bibnamefont
  {Denysenkov}}, \bibinfo {author} {\bibfnamefont {P.}~\bibnamefont {{van
  Os}}}, \ and\ \bibinfo {author} {\bibfnamefont {T.~F.}\ \bibnamefont
  {Prisner}},\ }\href {\doibase 10.1007/s00723-017-0930-9} {\bibfield
  {journal} {\bibinfo  {journal} {Applied Magnetic Resonance}\ }\textbf
  {\bibinfo {volume} {48}},\ \bibinfo {pages} {1263} (\bibinfo {year}
  {2017})}\BibitemShut {NoStop}%
\bibitem [{\citenamefont {Joshi}\ \emph {et~al.}(2016)\citenamefont {Joshi},
  \citenamefont {Miller}, \citenamefont {Ogden}, \citenamefont {Kavand},
  \citenamefont {Jamali}, \citenamefont {Ambal}, \citenamefont {Venkatesh},
  \citenamefont {Schurig}, \citenamefont {Malissa}, \citenamefont {Lupton},\
  and\ \citenamefont {Boehme}}]{joshiSeparatingHyperfineSpinorbit2016}%
  \BibitemOpen
  \bibfield  {author} {\bibinfo {author} {\bibfnamefont {G.}~\bibnamefont
  {Joshi}}, \bibinfo {author} {\bibfnamefont {R.}~\bibnamefont {Miller}},
  \bibinfo {author} {\bibfnamefont {L.}~\bibnamefont {Ogden}}, \bibinfo
  {author} {\bibfnamefont {M.}~\bibnamefont {Kavand}}, \bibinfo {author}
  {\bibfnamefont {S.}~\bibnamefont {Jamali}}, \bibinfo {author} {\bibfnamefont
  {K.}~\bibnamefont {Ambal}}, \bibinfo {author} {\bibfnamefont
  {S.}~\bibnamefont {Venkatesh}}, \bibinfo {author} {\bibfnamefont
  {D.}~\bibnamefont {Schurig}}, \bibinfo {author} {\bibfnamefont
  {H.}~\bibnamefont {Malissa}}, \bibinfo {author} {\bibfnamefont {J.~M.}\
  \bibnamefont {Lupton}}, \ and\ \bibinfo {author} {\bibfnamefont
  {C.}~\bibnamefont {Boehme}},\ }\href@noop {} {\bibfield  {journal} {\bibinfo
  {journal} {Appl. Phys. Lett.}\ }\textbf {\bibinfo {volume} {109}},\ \bibinfo
  {pages} {103303} (\bibinfo {year} {2016})}\BibitemShut {NoStop}%
\bibitem [{\citenamefont {Wedge}\ \emph {et~al.}(2012)\citenamefont {Wedge},
  \citenamefont {Timco}, \citenamefont {Spielberg}, \citenamefont {George},
  \citenamefont {Tuna}, \citenamefont {Rigby}, \citenamefont {McInnes},
  \citenamefont {Winpenny}, \citenamefont {Blundell},\ and\ \citenamefont
  {Ardavan}}]{wedgeChemicalEngineeringMolecular2012}%
  \BibitemOpen
  \bibfield  {author} {\bibinfo {author} {\bibfnamefont {C.~J.}\ \bibnamefont
  {Wedge}}, \bibinfo {author} {\bibfnamefont {G.~A.}\ \bibnamefont {Timco}},
  \bibinfo {author} {\bibfnamefont {E.~T.}\ \bibnamefont {Spielberg}}, \bibinfo
  {author} {\bibfnamefont {R.~E.}\ \bibnamefont {George}}, \bibinfo {author}
  {\bibfnamefont {F.}~\bibnamefont {Tuna}}, \bibinfo {author} {\bibfnamefont
  {S.}~\bibnamefont {Rigby}}, \bibinfo {author} {\bibfnamefont {E.~J.~L.}\
  \bibnamefont {McInnes}}, \bibinfo {author} {\bibfnamefont {R.~E.~P.}\
  \bibnamefont {Winpenny}}, \bibinfo {author} {\bibfnamefont {S.~J.}\
  \bibnamefont {Blundell}}, \ and\ \bibinfo {author} {\bibfnamefont
  {A.}~\bibnamefont {Ardavan}},\ }\href {\doibase
  10.1103/PhysRevLett.108.107204} {\bibfield  {journal} {\bibinfo  {journal}
  {Physical Review Letters}\ }\textbf {\bibinfo {volume} {108}} (\bibinfo
  {year} {2012}),\ 10.1103/PhysRevLett.108.107204}\BibitemShut {NoStop}%
\bibitem [{\citenamefont {Wolfowicz}\ \emph {et~al.}(2013)\citenamefont
  {Wolfowicz}, \citenamefont {Tyryshkin}, \citenamefont {George}, \citenamefont
  {Riemann}, \citenamefont {Abrosimov}, \citenamefont {Becker}, \citenamefont
  {Pohl}, \citenamefont {Thewalt}, \citenamefont {Lyon},\ and\ \citenamefont
  {Morton}}]{wolfowiczAtomicClockTransitions2013}%
  \BibitemOpen
  \bibfield  {author} {\bibinfo {author} {\bibfnamefont {G.}~\bibnamefont
  {Wolfowicz}}, \bibinfo {author} {\bibfnamefont {A.~M.}\ \bibnamefont
  {Tyryshkin}}, \bibinfo {author} {\bibfnamefont {R.~E.}\ \bibnamefont
  {George}}, \bibinfo {author} {\bibfnamefont {H.}~\bibnamefont {Riemann}},
  \bibinfo {author} {\bibfnamefont {N.~V.}\ \bibnamefont {Abrosimov}}, \bibinfo
  {author} {\bibfnamefont {P.}~\bibnamefont {Becker}}, \bibinfo {author}
  {\bibfnamefont {H.-J.}\ \bibnamefont {Pohl}}, \bibinfo {author}
  {\bibfnamefont {M.~L.~W.}\ \bibnamefont {Thewalt}}, \bibinfo {author}
  {\bibfnamefont {S.~A.}\ \bibnamefont {Lyon}}, \ and\ \bibinfo {author}
  {\bibfnamefont {J.~J.~L.}\ \bibnamefont {Morton}},\ }\href {\doibase
  10.1038/nnano.2013.117} {\bibfield  {journal} {\bibinfo  {journal} {Nature
  Nanotechnology}\ }\textbf {\bibinfo {volume} {8}},\ \bibinfo {pages} {561}
  (\bibinfo {year} {2013})}\BibitemShut {NoStop}%
\bibitem [{\citenamefont {Shiddiq}\ \emph {et~al.}(2016)\citenamefont
  {Shiddiq}, \citenamefont {Komijani}, \citenamefont {Duan}, \citenamefont
  {{Gaita-Ari{\~n}o}}, \citenamefont {Coronado},\ and\ \citenamefont
  {Hill}}]{shiddiqEnhancingCoherenceMolecular2016}%
  \BibitemOpen
  \bibfield  {author} {\bibinfo {author} {\bibfnamefont {M.}~\bibnamefont
  {Shiddiq}}, \bibinfo {author} {\bibfnamefont {D.}~\bibnamefont {Komijani}},
  \bibinfo {author} {\bibfnamefont {Y.}~\bibnamefont {Duan}}, \bibinfo {author}
  {\bibfnamefont {A.}~\bibnamefont {{Gaita-Ari{\~n}o}}}, \bibinfo {author}
  {\bibfnamefont {E.}~\bibnamefont {Coronado}}, \ and\ \bibinfo {author}
  {\bibfnamefont {S.}~\bibnamefont {Hill}},\ }\href {\doibase
  10.1038/nature16984} {\bibfield  {journal} {\bibinfo  {journal} {Nature}\
  }\textbf {\bibinfo {volume} {531}},\ \bibinfo {pages} {348} (\bibinfo {year}
  {2016})}\BibitemShut {NoStop}%
\bibitem [{\citenamefont {Collett}\ \emph {et~al.}(2019)\citenamefont
  {Collett}, \citenamefont {Ellers}, \citenamefont {Russo}, \citenamefont
  {Kittilstved}, \citenamefont {Timco}, \citenamefont {Winpenny},\ and\
  \citenamefont {Friedman}}]{collettClockTransitionCr7Mn2019}%
  \BibitemOpen
  \bibfield  {author} {\bibinfo {author} {\bibfnamefont {C.~A.}\ \bibnamefont
  {Collett}}, \bibinfo {author} {\bibfnamefont {K.-I.}\ \bibnamefont {Ellers}},
  \bibinfo {author} {\bibfnamefont {N.}~\bibnamefont {Russo}}, \bibinfo
  {author} {\bibfnamefont {K.~R.}\ \bibnamefont {Kittilstved}}, \bibinfo
  {author} {\bibfnamefont {G.~A.}\ \bibnamefont {Timco}}, \bibinfo {author}
  {\bibfnamefont {R.~E.~P.}\ \bibnamefont {Winpenny}}, \ and\ \bibinfo {author}
  {\bibfnamefont {J.~R.}\ \bibnamefont {Friedman}},\ }\href {\doibase
  10.3390/magnetochemistry5010004} {\bibfield  {journal} {\bibinfo  {journal}
  {Magnetochemistry}\ }\textbf {\bibinfo {volume} {5}},\ \bibinfo {pages} {4}
  (\bibinfo {year} {2019})}\BibitemShut {NoStop}%
\bibitem [{\citenamefont {Yang}\ \emph {et~al.}(2006)\citenamefont {Yang},
  \citenamefont {Wernsdorfer}, \citenamefont {Zakharov}, \citenamefont
  {Karaki}, \citenamefont {Yamaguchi}, \citenamefont {Isidro}, \citenamefont
  {Lu}, \citenamefont {Wilson}, \citenamefont {Rheingold}, \citenamefont
  {Ishimoto},\ and\ \citenamefont
  {Hendrickson}}]{yangFastMagnetizationTunneling2006a}%
  \BibitemOpen
  \bibfield  {author} {\bibinfo {author} {\bibfnamefont {E.-C.}\ \bibnamefont
  {Yang}}, \bibinfo {author} {\bibfnamefont {W.}~\bibnamefont {Wernsdorfer}},
  \bibinfo {author} {\bibfnamefont {L.~N.}\ \bibnamefont {Zakharov}}, \bibinfo
  {author} {\bibfnamefont {Y.}~\bibnamefont {Karaki}}, \bibinfo {author}
  {\bibfnamefont {A.}~\bibnamefont {Yamaguchi}}, \bibinfo {author}
  {\bibfnamefont {R.~M.}\ \bibnamefont {Isidro}}, \bibinfo {author}
  {\bibfnamefont {G.-D.}\ \bibnamefont {Lu}}, \bibinfo {author} {\bibfnamefont
  {S.~A.}\ \bibnamefont {Wilson}}, \bibinfo {author} {\bibfnamefont {A.~L.}\
  \bibnamefont {Rheingold}}, \bibinfo {author} {\bibfnamefont {H.}~\bibnamefont
  {Ishimoto}}, \ and\ \bibinfo {author} {\bibfnamefont {D.~N.}\ \bibnamefont
  {Hendrickson}},\ }\href {\doibase 10.1021/ic050093r} {\bibfield  {journal}
  {\bibinfo  {journal} {Inorganic Chemistry}\ }\textbf {\bibinfo {volume}
  {45}},\ \bibinfo {pages} {529} (\bibinfo {year} {2006})}\BibitemShut
  {NoStop}%
\bibitem [{\citenamefont {Collett}, \citenamefont {All{\~a}o~Cassaro},\ and\
  \citenamefont {Friedman}(2016)}]{collettPrecisionESRMeasurements2016}%
  \BibitemOpen
  \bibfield  {author} {\bibinfo {author} {\bibfnamefont {C.~A.}\ \bibnamefont
  {Collett}}, \bibinfo {author} {\bibfnamefont {R.~A.}\ \bibnamefont
  {All{\~a}o~Cassaro}}, \ and\ \bibinfo {author} {\bibfnamefont {J.~R.}\
  \bibnamefont {Friedman}},\ }\href {\doibase 10.1103/PhysRevB.94.220402}
  {\bibfield  {journal} {\bibinfo  {journal} {Physical Review B}\ }\textbf
  {\bibinfo {volume} {94}} (\bibinfo {year} {2016}),\
  10.1103/PhysRevB.94.220402}\BibitemShut {NoStop}%
\bibitem [{\citenamefont {Friedman}\ and\ \citenamefont
  {Sarachik}(2010)}]{friedmanSingleMoleculeNanomagnets2010}%
  \BibitemOpen
  \bibfield  {author} {\bibinfo {author} {\bibfnamefont {J.~R.}\ \bibnamefont
  {Friedman}}\ and\ \bibinfo {author} {\bibfnamefont {M.~P.}\ \bibnamefont
  {Sarachik}},\ }\href {\doibase 10.1146/annurev-conmatphys-070909-104053}
  {\bibfield  {journal} {\bibinfo  {journal} {Annual Review of Condensed Matter
  Physics}\ }\textbf {\bibinfo {volume} {1}},\ \bibinfo {pages} {109} (\bibinfo
  {year} {2010})}\BibitemShut {NoStop}%
\bibitem [{\citenamefont {Lawrence}\ \emph {et~al.}(2008)\citenamefont
  {Lawrence}, \citenamefont {Yang}, \citenamefont {Edwards}, \citenamefont
  {Olmstead}, \citenamefont {Ramsey}, \citenamefont {Dalal}, \citenamefont
  {Gantzel}, \citenamefont {Hill},\ and\ \citenamefont
  {Hendrickson}}]{lawrenceDisorderIntermolecularInteractions2008}%
  \BibitemOpen
  \bibfield  {author} {\bibinfo {author} {\bibfnamefont {J.}~\bibnamefont
  {Lawrence}}, \bibinfo {author} {\bibfnamefont {E.-C.}\ \bibnamefont {Yang}},
  \bibinfo {author} {\bibfnamefont {R.}~\bibnamefont {Edwards}}, \bibinfo
  {author} {\bibfnamefont {M.~M.}\ \bibnamefont {Olmstead}}, \bibinfo {author}
  {\bibfnamefont {C.}~\bibnamefont {Ramsey}}, \bibinfo {author} {\bibfnamefont
  {N.~S.}\ \bibnamefont {Dalal}}, \bibinfo {author} {\bibfnamefont {P.~K.}\
  \bibnamefont {Gantzel}}, \bibinfo {author} {\bibfnamefont {S.}~\bibnamefont
  {Hill}}, \ and\ \bibinfo {author} {\bibfnamefont {D.~N.}\ \bibnamefont
  {Hendrickson}},\ }\href {\doibase 10.1021/ic701416w} {\bibfield  {journal}
  {\bibinfo  {journal} {Inorganic Chemistry}\ }\textbf {\bibinfo {volume}
  {47}},\ \bibinfo {pages} {1965} (\bibinfo {year} {2008})}\BibitemShut
  {NoStop}%
\bibitem [{\citenamefont {Chen}\ \emph {et~al.}(2016)\citenamefont {Chen},
  \citenamefont {Ashkezari}, \citenamefont {Collett}, \citenamefont
  {All{\~a}o~Cassaro}, \citenamefont {Troiani}, \citenamefont {Lahti},\ and\
  \citenamefont {Friedman}}]{chenObservationTunnelingAssistedHighly2016}%
  \BibitemOpen
  \bibfield  {author} {\bibinfo {author} {\bibfnamefont {Y.}~\bibnamefont
  {Chen}}, \bibinfo {author} {\bibfnamefont {M.~D.}\ \bibnamefont {Ashkezari}},
  \bibinfo {author} {\bibfnamefont {C.~A.}\ \bibnamefont {Collett}}, \bibinfo
  {author} {\bibfnamefont {R.~A.}\ \bibnamefont {All{\~a}o~Cassaro}}, \bibinfo
  {author} {\bibfnamefont {F.}~\bibnamefont {Troiani}}, \bibinfo {author}
  {\bibfnamefont {P.~M.}\ \bibnamefont {Lahti}}, \ and\ \bibinfo {author}
  {\bibfnamefont {J.~R.}\ \bibnamefont {Friedman}},\ }\href {\doibase
  10.1103/PhysRevLett.117.187202} {\bibfield  {journal} {\bibinfo  {journal}
  {Physical Review Letters}\ }\textbf {\bibinfo {volume} {117}} (\bibinfo
  {year} {2016}),\ 10.1103/PhysRevLett.117.187202}\BibitemShut {NoStop}%
\end{thebibliography}%

\end{document}